\documentclass[aps,amsmath,amssymb,prd,showpacs,floatfix,preprint,superscriptaddress,nofootinbib,12pt]{article}
\usepackage{jheppub}
\usepackage{bm}
\usepackage{ulem}

\usepackage{axodraw4j}
\usepackage{pstricks}
\usepackage{pgfplots}
\usepackage{amsmath}
\usepackage{color}

\usepackage{epstopdf}

\allowdisplaybreaks[3]

\begin{document}

\preprint{}

\title{
\begin{center}
Large $N$ Thirring Matter in Three Dimensions
\end{center}
}

\author[]{Mikhail Goykhman}

\affiliation[]{Enrico Fermi Institute, University of Chicago,
5620 S. Ellis Av., Chicago, IL 60637, USA}
\emailAdd{goykhman@uchicago.edu}

\abstract{
In this paper we calculate  properties of
the three-dimensional system of $N$ species of fermions at zero temperature and finite chemical potential, with the
four-fermionic interaction of the Thirring type.
We observe that this model fits consistently into framework of
the Landau Fermi liquid theory, and possesses a non-trivial zeroth and first Landau parameters.
Our result is derived to all orders of the Thirring coupling constant and to the leading order of
the large-$N$ expansion. In particular we solve for the exact current-current
correlation function, and show that it exhibits a singular behavior
at zero frequency and twice of the Fermi momentum.}

\maketitle

\section{Introduction}\label{sec:intro}

Consider theory of $N$ species of fermions $\psi_i$, $i=1,\dots,N$ in the $d$-dimensional space-time.
A straightforward
way to make this theory dynamical is to switch on a local four-fermionic interaction of the Gross-Neveu $(\bar\psi\psi)^2$ or the Thirring $(\bar\psi\gamma^\mu\psi)^2$ form \cite{Gross:1974jv,Thirring:1958in}.
At weak interaction the dimension of the four-fermionic coupling constant
is $2-d$. Therefore
such an interaction is power-counting non-renormalizable when $d>2$.
However the three-dimensional models with a local four-fermionic interaction
of the Gross-Neveu and Thirring form
are known to be renormalizable and solvable
in the large $N$ expansion
\cite{Parisi:1975,Rosenstein:1988pt,Rosenstein:1988dj,Rosenstein:1990nm,Gomez:1990nm,Hands:1994kb}.
Such three-dimensional interacting models therefore represent a rather simple exactly solvable fermionic systems
which exhibit a non-trivial dynamics.

In this paper we want to discuss a three-dimensional interacting fermionic
matter at finite density. The specific model which we choose is
the three-dimensional large $N$
Thirring model taken at finite chemical potential for the $U(1)$
particle number.
We will solve this model to the leading order in the $1/N$ expansion,
and to all orders of the Thirring coupling constant.
We will refrain our
consideration mostly to the case of zero temperature, and we will set the bare fermionic mass to be zero.
In the discussion section we outline the future work and possible interesting generalizations
and extensions which can be performed.

One of the phases in which one can
find a quantum dense fermionic system is described by the Landau Fermi liquid theory
\cite{baym2008landau,abrikosov1975methods,baym1976landau}. This is
a finite-density non-condensed state exhibiting a long-lived fermionic
quasiparticle excitations.
It is defined in the low-temperature regime $T/\mu\ll 1$
in which quasiparticle excitations predominantly exist near the Fermi surface.
Interaction of such quasiparticles is characterized by the coupling
constants known as the Landau parameters.
It has recently been demonstrated that the large-$N$
Chern-Simons-fermion theory provides an example of a microscopic realization
of a non-trivial Landau Fermi liquid state \cite{Geracie:2015drf}.
\footnote{Literature on the Chern-Simons-matter theories includes \cite{Giombi:2011kc,Aharony:2011jz,Aharony:2012nh,GurAri:2012is,Aharony:2012ns,
Yokoyama:2012fa,Jain:2013py,Jain:2013gza,Takimi:2013zca,Jain:2014nza,
Moshe:2014bja,Inbasekar:2015tsa,Aharony:2015mjs,Yokoyama:2016sbx,Gur-Ari:2016xff}.}
Holographic non-Fermi liquid models
are known \cite{Lee:2008xf,Cubrovic:2009ye}.

In this paper we show that the dense large-$N$ Thirring model in the low-temperature regime behaves as the Landau
Fermi liquid.
Our argument is constructed analogously to \cite{Geracie:2015drf}.
First, we calculate the exact fermionic
propagator and the four-fermionic scattering amplitude, and use these
to calculate the Landau parameters microscopically.
Second, we use the Landau Fermi liquid theory expressions for
the inverse compressibility and the heat capacity to solve for the Landau parameters. This is possible to do since these thermodynamic quantities can be found by a separate calculation. We show that both these methods give the same values of the Landau parameters.

The relative simplicity of the Thirring model, as compared to the Chern-Simons-fermion
model of \cite{Geracie:2015drf}, allows one to straightforwardly calculate the current-current
correlation function at a finite value of the spatial momentum.
In the Fermi liquid this correlation function exhibits a singular behavior
at zero frequency and twice the Fermi momentum.
Experimentally this finite-momentum singularity manifests in a rippling pattern of the
charge screening response to insertion of an external charged impurity,
known as the Friedel oscillations.
A holographic example of a similar oscillatory screening mechanism
has recently been obtained in \cite{Blake:2014lva}.
We derive the current-current correlation function
for the large $N$ Thirring model and demonstrate explicitly
that it exhibits a singular behavior at zero frequency and twice the Fermi momentum. We show that the value of the Fermi momentum is consistent with the Fermi-Dirac quasiparticle distribution and the Luttinger's theorem.

The paper is organized as follows.
In section \ref{sec:LFL} we briefly summarize the key points of the Landau Fermi liquid theory.
In section \ref{sec:Thirring} we formulate the three-dimensional Thirring model at finite chemical potential,
and solve for the exact fermionic propagator. In particular this allows one
to determine the fermionic distribution function, and calculate the entropy 
and the heat capacity.
From the fermionic distribution function one can read off the value of the Fermi momentum.
In section \ref{sec:curcur} we calculate the current-current
correlation function. We show that it exhibits a zero-frequency singular behavior
at the value of the momentum, equal to twice the Fermi momentum found in section \ref{sec:Thirring}. In section \ref{sec:Landaupar} we calculate the values of the Landau
parameters from both the microscopic and the thermodynamic perspective,
and demonstrate an agreement of both methods. We discuss our results
and comment on further work directions in section \ref{sec:disc}.

\section{Landau Fermi liquid theory}\label{sec:LFL}

In this section we will briefly review some aspects of the Landau Fermi liquid theory.
We refer the reader to  \cite{baym2008landau,abrikosov1975methods,baym1976landau,Geracie:2015drf} for the detailed presentation, and in this section we merely outline the very basic
statements, in order to make the paper sufficiently self-contained.

The Landau Fermi liquid theory is a low-temperature $T/\mu\ll 1$
theory of a quantum fermionic liquid. Its underlying assumption is that
the fundamental low-energy degrees of freedom of the system are long-lived quasiparticles
in the vicinity of the Fermi surface.
In 2+1 dimensions, which is the case of interest of the present paper, there is
no spin degree of freedom, and each quasiparticle state is characterized
only by the value of its momentum ${\bf p}$. Define $\varepsilon ({\bf p})$
to be an energy of a quasiparticle, and $n({\bf p})$ to be an occupation number.

Interaction of a pair of quasiparticles
with momenta ${\bf p}$ and ${\bf p'}$ is described by a function $f({\bf p},{\bf p'})$,
which is introduced in the following way. Suppose occupation numbers of quasiparticles
receive a perturbation $\delta n({\bf p})$. In Landau Fermi liquid this results
in the following perturbation of the energy of a given quasiparticle:
\begin{equation}
\delta \varepsilon ({\bf p})=\int \frac{d^3{\bf p'}}{(2\pi)^3}\,
f({\bf p},{\bf p'})\,\delta n({\bf p'})\,.
\end{equation}
The Fermi velocity and the effective mass of a quasiparticle are defined as
\begin{equation}
v_F=\frac{\partial \varepsilon ({\bf p})}{\partial |{\bf p}|}\Bigg|_{|{\bf p}|=p_F}\,,
\quad\quad
m^\star =\frac{p_F}{v_F}\,.
\end{equation}

In this paper we will be considering $N$ fermionic flavors,
and therefore the observables have an extra index structure:
\begin{align}
\varepsilon ^i_{\;\; j}({\bf p})&=\delta ^i_j \,\varepsilon ({\bf p})\,,\quad
n ^i_{\;\; j}({\bf p})=\delta ^i_j n\, ({\bf p})\,,\\
f^{i\;\;\;\; l}_{\;\; j,\;\; k}({\bf p},{\bf p'})
&=f^{(d)}({\bf p},{\bf p'})\delta ^i_j\delta ^l_k+f^{(e)}
({\bf p},{\bf p'})\delta ^i_k\delta ^l_j
\end{align}
In the large-$N$ limit the dominant contribution comes
from the direct channel, $f^{(d)}$, while the exchange channel,
$f^{(e)}$, is suppressed \cite{Geracie:2015drf}. We therefore will be considering $f^{(d)}=f$. In the low-temperature
limit interacting qusiparticles are restricted to the vicinity
of the Fermi surface, $|{\bf p}|=|{\bf p'}|=p_F$.
We define $\theta$ to be an angle between ${\bf p}$ and ${\bf p'}$,
and expand 
\begin{equation}
\label{fexpansion}
f(\theta)=\frac{1}{\nu(\epsilon_F)}\left(F_0+2\sum_{n=1}
^\infty F_n\cos(n\theta)\right)\,.
\end{equation}
The coefficients of expansion, $F_n$, $n=0,1,\dots$,
are called the Landau parameters. We have defined the density
of states on the Fermi surface as
\begin{equation}
\nu(\epsilon_F)=\frac{Nm^\star}{2\pi}\,.	
\end{equation}
Using the framework outlined above one can derive the values
of various observables in terms of the Landau
parameters. First of all the quasiparticle effective mass
of a relativistic Landau Fermi liquid is given by
\begin{equation}
\label{LFLeffectivemass}
m^\star=\mu\, (1+F_1)\,.	
\end{equation}
Quasiparticles with the effective mass $m^\star$ obeying
the Fermi-Dirac distribution exhibit the following low-temperature behavior of the heat capacity
\begin{equation}
\label{LFLheatcapacity}
c=\frac{\pi}{6}\, N\, m^\star\, T\,.	
\end{equation}
We will also find useful the following expression for the inverse compressibility
\begin{equation}
\label{LFLinvsersecompressibility}
\kappa^{-1}=n^2\left(\frac{\partial\mu}{\partial n}\right)_T=
\frac{2\pi n^2}{Nm^\star}\,(1+F_0)\,.	
\end{equation}

Interacting quasiparticles can be described in the framework
of quantum field theory. The fermionic propagator, in the Lorentzian signature, near the Fermi surface acquires the form
\begin{equation}
\label{LFLpropagator}
G(p)=\frac{Z}{\omega-v_F(|{\bf p}|-p_F)+i\,\epsilon\, {\rm sgn}
(|{\bf p}|-p_F)}\,\frac{u\otimes \bar u}{u^\dagger u}\,,	
\end{equation}
where $u(p)$ is an on-shell spinor describing quasiparticle, and
$Z$ is the wave-function renormalization constant.
The Landau parameters are calculated as
\begin{equation}
\label{microscopicf}
f^{i\;\;\; k}_{\;\; j,\;\;\; l}(\theta)
=Z^2\lim _{q^0\rightarrow 0}\lim _{{\bf q}\rightarrow 0}	
V^{i\;\;\; k}_{\;\; j,\;\;\; l}(p,k,q)\,,
\end{equation}
where the on-shell four-fermionic vertex is
\begin{equation}
\label{1PIamplitude}
V^{i\;\;\; k}_{\;\; j,\;\;\; l}(p,k,q)=\frac{1}{(u^\dagger u)^2}
u_\alpha (p+q)u_\gamma(k)V^{\alpha\;\;\;\; \gamma}_{\;\;\; \delta,\;\;\;\; \beta}\;
{}^{i\;\;\; k}_{\;\; j,\;\;\; l}(p,k,q)\bar u^\beta (k+q)\bar u^\delta (p)\,,
\end{equation}
and the one-particle irreducible four-fermionic amplitude is given by
\begin{equation}
\label{1PIvertex}
	V^{\alpha\;\;\;\; \gamma}_{\;\;\; \delta\;\;\;\; \beta}\;
{}^{i\;\;\; k}_{\;\; j\;\;\; l}(p,k,q)=
\left\langle\bar\psi^{\alpha i}(-p-q)
\bar\psi ^{\gamma k}(-k)\psi_{\delta l}(p)
\psi _{\beta j}(k+q)\right\rangle_{1PI}\,.
\end{equation}

\section{Thirring model}\label{sec:Thirring}

In this paper we study the system of $N$ massless Dirac fermions
interacting via the four-fermionic Thirring coupling.
We will be working in the Euclidean space, with $x^{1,2}$ being the space coordinates,
and $x^3$ being the Euclidean time coordinate. We consider
the model at finite chemical potential $\mu$ for the $U(1)$ particle number current
$j^\mu=\bar\psi^n\gamma^\mu\psi_n$. We will mostly be considering the
system at zero temperature.
The Lagrangian is given by
\begin{equation}
\label{Lagrangian}
{\cal L}=\bar\psi^n\gamma^\mu \partial_\mu\psi_n+\mu\,\bar\psi^n \gamma^3\psi_n+\frac{1}{2\sigma^2N}\,
\bar\psi^n\gamma^\mu\psi_n\,\bar\psi^m\gamma_\mu\psi_m\,.
\end{equation}
Here $\sigma^2$ has a dimension of mass, and the case of free fermions corresponds
to taking the limit $\sigma^2\rightarrow\infty$. The conjugate spinor in the Euclidean space
is $\bar\psi=\psi^\dagger$. An extra prefactor of $1/2$
in (\ref{Lagrangian}) is introduced for further convenience.
In the case $\sigma^2>0$ the interaction is repulsive, while
in the case $\sigma^2<0$ the interaction is attractive. 

The model (\ref{Lagrangian}) is exactly solvable in the large-$N$
limit. In this paper we are interested in the solution at the leading order in the $1/N$ expansion.
Let us begin by solving for the exact fermionic propagator,
\begin{equation}
\langle \psi_m(p)\bar\psi ^n(-q)\rangle= \delta_m^n G(p)(2\pi)^3 \delta (p-q)\,,
\end{equation}
We will be looking for the solution of the form
\begin{equation}
G(p)=\frac{1}{i\tilde p_\mu\gamma^\mu +\Sigma (p)}\,,
\end{equation}
where $\tilde p_\mu=p_\mu -i\mu \delta_{\mu,3}$.
The fermionic self-energy $\Sigma (p)$ satisfies the Schwinger-Dyson equation,
\begin{equation}
\label{SDforSigma}
\Sigma(p)=\frac{1}{\sigma^2}\gamma^\mu\int\frac{d^3r}{(2\pi)^3}{\rm Tr} (G(r)\gamma_\mu)\,.
\end{equation}
It can be derived in the path integral framework, as
well as from the following diagramatic consideration:
\begin{center}
\fcolorbox{white}{white}{
 \scalebox{.7}{
  \begin{picture}(466,166) (47,-59)
    \SetWidth{1.0}
    \SetColor{Black}
    \Line[arrow,arrowpos=0.5,arrowlength=5,arrowwidth=5,arrowinset=1](48,-26)(128,-26)
    \GOval(160,-26)(32,32)(0){0.882}
    \Line[arrow,arrowpos=0.5,arrowlength=5,arrowwidth=5,arrowinset=1](48,-26)(128,-26)
    \Line[arrow,arrowpos=0.5,arrowlength=5,arrowwidth=5,arrowinset=1](192,-26)(272,-26)
    \Text(155,-30)[lb]{\Large{\Black{$\Sigma$}}}
    \Text(305,-30)[lb]{\Large{\Black{$=$}}}
    %
    \SetWidth{1.0}
    \Line[arrow,arrowpos=0.5,arrowlength=5,arrowwidth=5,arrowinset=1](48,-26)(128,-26)
    \Line[arrow,arrowpos=0.5,arrowlength=5,arrowwidth=5,arrowinset=1](352,-26)(432,-26)
    \Line[arrow,arrowpos=0.5,arrowlength=5,arrowwidth=5,arrowinset=1](432,-26)(512,-26)
    \Arc[arrow,arrowpos=0.5,arrowlength=5,arrowwidth=5,arrowinset=1,clock](433.6,23.6)(49.626,-91.848,-249.228)
    \Arc[arrow,arrowpos=0.5,arrowlength=5,arrowwidth=5,arrowinset=1,clock](430.4,23.6)(49.626,69.228,-88.152)
    \Vertex(432,-26){6}
    \GOval(432,70)(32,32)(0){0.882}
    \Text(428,66)[lb]{\Large{\Black{$G$}}}
  
  \end{picture}
  }
}
\end{center}

It is clear from equation (\ref{SDforSigma}) that the solution is momentum independent,
$\Sigma(p)\equiv {\rm const}$, which is the consequence of
a contact nature of the Thirring interaction.
After regularizing the integrals over $r_3$ and $r_s=\sqrt{r_1^2+r_2^2}$, we arrived at
\begin{align}
\Sigma_1&=0\,,\quad \quad \Sigma_2=0\,,\quad\quad
\Sigma_3=-\frac{\hat\mu^2}{4\pi\sigma^2}\,,\label{Sigma3}
\end{align}
where
\begin{equation}
\label{hatmudefinition}
\hat\mu=\mu +\Sigma_3\,.
\end{equation}
Therefore the exact fermionic propagator is the same
as the free fermionic propagator, but with the renormalized chemical potential
(\ref{hatmudefinition}),
\begin{align}
G(p)=\frac{1}{i\hat p_\mu \gamma^\mu}\,,\quad\quad
\hat p_\mu &=p_\mu -i\hat \mu \delta _{\mu,3}\,.
\end{align}

From the exact fermionic propagator one can derive the distribution
function for fermionic states in the momentum space (quasiparticle occupation number),
\begin{equation}
n({\bf p})=\int\frac{dp_3}{2\pi}\int\frac{d^3q}{(2\pi)^3}
\left\langle\bar\psi ^n (q)\gamma^3\psi _n(p)\right\rangle =N\,\theta (\hat \mu -|{\bf p}|)\,.
\end{equation}
This distribution function indicates a presence of the Fermi
surface, with the Fermi momentum given by
\begin{equation}
p_F=\hat\mu\,.\label{pfThirring}
\end{equation}
Knowing the distribution function one can determine
the density of fermions
\begin{equation}
\label{fermiondensity}
n=-\frac{\partial \log Z}{\partial\mu}=\int \frac{d^2p}{(2\pi)^2}n ({\bf p})=N\,\frac{\hat\mu^2}{4\pi}\,.
\end{equation}
Notice that the expression (\ref{fermiondensity})
is derived without explicit use of expression for the free energy,
and is a manifestation of the Luttinger's theorem
for fermions at zero temperature, with the Fermi momentum $p_F=\hat\mu$.

We have therefore demonstrated that zero-temperature occupation number
for fermionic states is given by a step function, and we have
identified the Fermi momentum by the location of the step. In the next section we demonstrate how a singular
structure at zero frequency and twice of the Fermi momentum is manifested in the
current-current correlation function.

In this paper we are mostly interested in the case
of zero temperature. However one of the hallmarks
of the Landau Fermi liquid is a linear temperature
dependence of the low-temperature heat capacity (\ref{LFLheatcapacity}). Therefore in the remaining part of this section we briefly outline derivation of the
the fermionic propagator and the occupation number at finite temperature.

The Schwinger-Dyson equation at finite temperature is
\footnote{Notice that a finite-temperature calculation in the
Thirring theory is simpler than in the Chern-Simons-matter theories,
because the absence of gauge symmetry means that one does not
have to keep track of the holonomy of the gauge field
on the temporal circle. The importance of the holonomy in the thermal
Chern-Simons-matter theories was first pointed out in \cite{Aharony:2012ns}.}
\begin{equation}
\Sigma(p)=\frac{1}{\sigma^2}\gamma^\mu\,\frac{1}{\beta}\,\sum_n\,\int\frac{d^2r}{(2\pi)^2}{\rm Tr} (G(r)\gamma_\mu)\Bigg|_{\tilde r_3=2\pi(n+1/2)/\beta-i\mu}\,.
\end{equation}
The solution for $\Sigma_{1,2}$ is again trivial, while $\Sigma_3$
is a constant, shifting the chemical potential, $\hat\mu=\mu+\Sigma_3$.
Regularizing the sum over $n$, and the integral over $r_s$ one can obtain the finite-temperature equation for $\Sigma_3$.
The precise form of this equation is not essential for our present purposes and will be derived elsewhere.

The occupation number can again be calculated
from the fermionic propagator,
\begin{align}
n({\bf p})&=\frac{1}{\beta}\,\sum_n\,{\rm Tr}\left(\gamma^3 G\left(p_3=\frac{2\pi\left(n+\frac{1}{2}\right)}{\beta}-i\mu,{\bf p}\right)\right)\\
&=\frac{1}{2}\left(\tanh \left(\frac{1}{2}\beta (p_s+\hat\mu)\right)-
\tanh \left(\frac{1}{2}\beta (p_s-\hat\mu)\right)\right)\,,
\end{align}
which is the Fermi-Dirac distribution of massless Dirac fermion with energy $E=p_s$ at the chemical potential $\hat\mu$.
Knowing the distribution function one can determine
the entropy and the heat capacity of the system,
\begin{equation}
\label{heatcapacityThirring}
s=\frac{\pi}{6}\,N\,\hat\mu\, T\,,\qquad c=\frac{\pi}{6}\,N\,\hat\mu \,T\,.
\end{equation}
This agrees with the Landau Fermi liquid expression (\ref{LFLheatcapacity}), provided the quasiparticle effective
mass is $m^\star=\hat\mu$.

\section{Current-current correlator}\label{sec:curcur}

In the previous section we showed that fermionic
distribution function of the zero-temperature Thirring
model at finite chemical potential $\mu$ is a step function,
with the step located at the momentum (\ref{pfThirring}). This is what we expect to see when
a sharp Fermi surface is formed. In this section we want to
provide an independent verification of existence of the Fermi surface.

One can detect
the Fermi surface experimentally by observing response of the
system to an external charged impurity. Due to the Fermi surface
the charge screening mechanism will exhibit a rippling pattern,
which can be traced back to a singular structure of the current-current
correlation function at zero frequency and finite momentum.
In this section we perform a calculation of the current-current
correlation function in the Thirring model, and show that
it is singular at zero frequency and the momentum equal to
twice of the Fermi momentum (\ref{pfThirring}), as it is expected in Fermi liquids.

Consider the global $U(1)$ current, $j^\mu (x)=\bar\psi^m(x)\gamma^\mu\psi_m(x)$.
Corresponding bi-fermionic vertex is
\begin{equation}
(T^\mu)^{\alpha}_{\;\;\beta}(p,q)\,\delta^n_m=\langle \bar\psi_m^\alpha (p+q)
\psi_\beta^n(p)j^\mu (-q)\rangle\,.
\end{equation}
We propose the following ansatz to incorporate
the spinor structure of the vertex
\begin{equation}
T^\mu=T^\mu_{\;\;\nu}\gamma^\nu\,.
\end{equation}
This vertex satisfies the Schwinger-Dyson equation
\begin{equation}
T^\mu_{\;\;\nu} (q)\gamma^\nu =\gamma^\mu
+\frac{1}{\sigma^2}\gamma^\nu\, T^{\mu\lambda}(q)
\int\frac{d^3r}{(2\pi)^3}\,{\rm Tr}\, (\gamma_\nu G(r+q)\gamma_\lambda G(r))\,,
\end{equation}
which diagramatically is depicted as (the large black vertex stands for four-fermionic coupling, the small black vertex stands for free contraction point, the internal fermionic lines are full propagators)
\begin{center}
\fcolorbox{white}{white}{
\scalebox{.7}{
  \begin{picture}(522,168) (13,-3)
    \SetWidth{1.0}
    \SetColor{Black}
    \GOval(96,72)(16,16)(0){0.882}
    \Line[arrow,arrowpos=0.5,arrowlength=5,arrowwidth=5,arrowinset=1](24,136)(80,80)
    \Line[arrow,arrowpos=0.5,arrowlength=5,arrowwidth=5,arrowinset=1](80,64)(24,8)
    \Text(180,65)[lb]{\Large{\Black{$=$}}}
    \Line[arrow,arrowpos=0.5,arrowlength=5,arrowwidth=5,arrowinset=1](168,136)(230,70)
    \Line[arrow,arrowpos=0.5,arrowlength=5,arrowwidth=5,arrowinset=1](230,70)(168,8)
    \SetWidth{0.0}
    \Vertex(230,72){2.828}
    \Text(296,65)[lb]{\Large{\Black{$+$}}}
    \SetWidth{1.0}
    \GOval(440,72)(16,16)(0){0.882}
    \Arc[arrow,arrowpos=0.5,arrowlength=5,arrowwidth=5,arrowinset=1](391.605,88.734)(48.4,-0.869,190.397)
    \Arc[arrow,arrowpos=0.5,arrowlength=5,arrowwidth=5,arrowinset=1](391.682,56.178)(48.319,170.684,359.789)
    \SetWidth{0.0}
    \Vertex(344,72){8.485}
    \SetWidth{1.0}
    \Arc[arrow,arrowpos=0.5,arrowlength=5,arrowwidth=5,arrowinset=1,clock](283.22,65.488)(61.128,6.116,-70.126)
    \Text(55,115)[lb]{\Large{\Black{$p+q$}}}
    \Text(62,22)[lb]{\Large{\Black{$p$}}}
    \Text(10,140)[lb]{\Large{\Black{$\alpha$}}}
    \Text(156,140)[lb]{\Large{\Black{$\alpha$}}}
    \Text(291,140)[lb]{\Large{\Black{$\alpha$}}}
    \Text(10,-8)[lb]{\Large{\Black{$\beta$}}}
    \Text(156,-8)[lb]{\Large{\Black{$\beta$}}}
    \Text(291,-8)[lb]{\Large{\Black{$\beta$}}}
    \Photon(112,72)(152,72){7.5}{2}
    \Photon(232,72)(272,72){7.5}{2}
    \Photon(456,72)(496,72){7.5}{2}
    \Text(374,144)[lb]{\Large{\Black{$r+q$}}}
    \Text(390,-8)[lb]{\Large{\Black{$r$}}}
    \Text(130,84)[lb]{\Large{\Black{$q$}}}
    \Text(155,66)[lb]{\Large{\Black{$\mu$}}}
    \Text(276,66)[lb]{\Large{\Black{$\mu$}}}
    \Text(500,66)[lb]{\Large{\Black{$\mu$}}}
    \Arc[arrow,arrowpos=0.5,arrowlength=5,arrowwidth=5,arrowinset=1,clock](276.668,72.96)(67.397,67.002,-2.517)
  \end{picture}
  }
}
\end{center}

Here we have noticed that once again the momentum dependence is only on the
total momentum $q$. This in turn leaves us with just an algebraic equation
\begin{equation}
T^\mu_{\;\;\nu}=\delta^\mu_{\;\;\nu}-\frac{1}{\sigma^2}T^{\mu}_{\;\;\lambda}v^\lambda_{\;\;\nu}\,,
\end{equation}
where we have defined
\begin{equation}
\label{vmunudefinition}
v_{\mu\nu}={\rm Tr}(\gamma_\mu\gamma_\alpha\gamma_\nu
\gamma_\beta)\,\int\frac{d^3r}{(2\pi)^3}\frac{1}{\hat r^2(\hat r+q)^2}\hat r^\alpha (
\hat r+q)^\beta\,.
\end{equation}

The solution is
\begin{equation}
T=\sigma^2(\sigma^2 \, I+v)^{-1}\,.
\end{equation}
The current-current correlator can then be straightforwardly calculated,
\begin{align}
\label{currentcurrent}
\langle j^\nu (-q)j^\mu (q)\rangle &=
N\,T^\mu_{\;\;\lambda}(q)\int\frac{d^3r}{(2\pi)^3}\,{\rm Tr}\, (\gamma^\nu G(r+q)\gamma^\lambda G(r))\\
&=-(T\, v)^{\mu\nu}\\
&=-N\,\sigma^2((\sigma^2 \, I+v)^{-1}\, v)^{\mu\nu}\label{currentcorrelator}\,.
\end{align}

\begin{figure}
	\centering
\scalebox{0.8}{
		\includegraphics[width=1\textwidth]{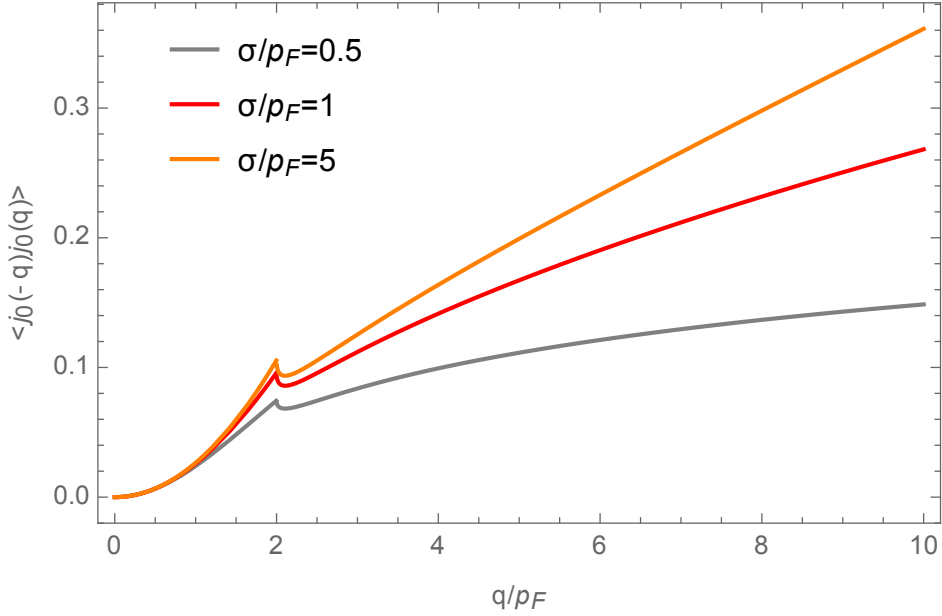}
		}
			\caption{The real part of the density-density correlation function $\left\langle j_0(0,-q)j_0(0,q)\right\rangle$, at various values of the coupling constant $\sigma$. The imaginary part identically vanishes at zero frequency.}
	\label{fig:densitycorrelation}
	\end{figure}

It requires some work to calculate $v_{\mu\nu}$ at non-zero
momentum, however when $q_3=0$ the derivation somewhat simplifies, but still remains cumbersome.
We provide the details of its calculation in Appendix \ref{App:A}.
The most important feature is the singular structure of the correlation function
at $q_s=2p_F=2\hat\mu$. We plot the result at various
values of the coupling constant in figure \ref{fig:densitycorrelation}.
At finite value of the coupling $\sigma$ we obtain the asymptotic expression
(remember $\sigma^2$ has dimension of mass)
\begin{equation}
\frac{1}{N}\,	\left\langle j_0(0,-{\bf q})j_0(0,{\bf q})\right\rangle\simeq \sigma ^2-\frac{32 \sigma ^4}{q_s}+{\cal O}\left(1/q_s^2\right)\,.
\end{equation}
On the other hand if the free limit $\sigma\rightarrow\infty$
is taken first, then the asymptotic behavior is 
\begin{equation}
\frac{1}{N}\,	\left\langle j_0(0,-{\bf q})j_0(0,{\bf q})\right\rangle_{\sigma\rightarrow \infty}\simeq \frac{q_s}{32}+\frac{\mu}{16}+{\cal O}\left(1/q_s\right)\,.
\end{equation}

We also calculate
\begin{equation}
\frac{1}{N}\,\lim_{q_s\rightarrow 0}\left\langle j_{\parallel} (0,-{\bf q})j_\parallel (0,{\bf q})\right\rangle=	
\frac{p_F^2}{4\pi\mu}
\end{equation}
which then can be substituted into the Kubo formula to give the following expression for the AC conductivity at low frequency
\begin{equation}
\sigma(\omega)=	N\,\frac{p_F^2}{4\pi\mu}\frac{1}{-i\omega}\,.
\end{equation}
This agrees with the Drude model, with infinitely long-lived charge carriers
with the density given by (\ref{fermiondensity}).

\section{Landau parameters}\label{sec:Landaupar}

In the Landau Fermi liquid theory the interaction of quasiparticles on the Fermi surface
can be described by the Landau parameters, as we briefly
reviewed in section \ref{sec:LFL}.
In the first part of this section we derive the Landau parameters
for the Thirring model from microscopic considerations.
We begin by solving the Schwinger-Dyson equation for the
fermionic four-point function (\ref{1PIvertex}). Knowing the on-shell fermionic
states on the Fermi surface, we subsequently derive the amplitude (\ref{1PIamplitude}), and the 
quasiparticle interaction function (\ref{microscopicf}).

Thermodynamic observables in the Landau Fermi liquid theory
can be expressed in terms of the Landau parameters.
As reviewed in section \ref{sec:LFL}, the heat capacity
and the inverse compressibility can be calculated using
(\ref{LFLheatcapacity}), (\ref{LFLinvsersecompressibility}).
The reverse of this procedure expresses the Landau parameters
in terms of the known thermodynamic observables.
In the second part of this section we show that such a method
gives the values of the Landau parameters agreeing with the
microscopic result.

\subsection{Four-fermionic vertex}
Consider the four-fermionic one-particle irreducible amplitude
in the direct channel
\begin{equation}
V^{\alpha\;\;\;\;\gamma}_{\;\;\delta\;\;,\;\;\beta}(q)\,
\delta^i_l\,\delta ^j_k=
\left\langle\bar\psi^{\alpha i}(-p-q)\bar\psi ^{\gamma k}(-k)
\psi_{\delta l}(p)\psi _{\beta j}(k+q)\right\rangle _{1PI}
\end{equation}
It satisfies the Schwinger-Dyson equation, which in the large-$N$
limit is written as
\begin{align}
\label{SD:vertex}
V^{\alpha\;\;\;\;\gamma}_{\;\;\delta\;\;,\;\;\beta}(q)=\frac{1}{\sigma^2}(\gamma^\mu)^\alpha_{\;\;\delta}
(\gamma_\mu)^\gamma_{\;\;\beta}+\frac{1}{\sigma^2}(\gamma^\mu)^\alpha_{\;\;\delta}
\int\frac{d^3r}{(2\pi)^3}{\rm Tr}_1\left(G(r)\gamma_\mu G(r+q)
V^{\;\;\;\;\gamma}_{\;\;\;\;,\;\;\beta}(q)\right)
\end{align}
The pairs of indices $(\alpha,\delta)$ and $(\gamma,\beta)$ are separated by comma
in the notation for the vertex, and belong to different terms in the direct product of spinor structure.
Diagramatically the Schwinger-Dyson equation (\ref{SD:vertex})
looks like
\begin{center}
\fcolorbox{white}{white}{
 \scalebox{.7}{
  \begin{picture}(566,118) (11,-27)
    \SetWidth{1.0}
    \SetColor{Black}
    \GOval(96,22)(48,16)(0){0.882}
    \Line[arrow,arrowpos=0.5,arrowlength=5,arrowwidth=5,arrowinset=1](16,54)(80,54)
    \Line[arrow,arrowpos=0.5,arrowlength=5,arrowwidth=5,arrowinset=1](112,54)(176,54)
    \Text(206,20)[lb]{\Large{\Black{$=$}}}
    \Vertex(288,22){8}
    \Arc[arrow,arrowpos=0.5,arrowlength=5,arrowwidth=5,arrowinset=1,clock](233.333,52.667)(62.681,-29.291,-91.219)
    \Line[arrow,arrowpos=0.5,arrowlength=5,arrowwidth=5,arrowinset=1](80,-10)(16,-10)
    \Line[arrow,arrowpos=0.5,arrowlength=5,arrowwidth=5,arrowinset=1](176,-10)(112,-10)
    \Arc[arrow,arrowpos=0.5,arrowlength=5,arrowwidth=5,arrowinset=1,clock](233.333,-8.667)(62.681,91.219,29.291)
    \Text(357,18)[lb]{\Large{\Black{$+$}}}
    \Arc[arrow,arrowpos=0.5,arrowlength=5,arrowwidth=5,arrowinset=1,clock](342.667,-8.667)(62.681,150.709,88.781)
    \Arc[arrow,arrowpos=0.5,arrowlength=5,arrowwidth=5,arrowinset=1,clock](377.333,-8.667)(62.681,91.219,29.291)
    \Arc[arrow,arrowpos=0.5,arrowlength=5,arrowwidth=5,arrowinset=1,clock](486.667,-8.667)(62.681,150.709,88.781)
    \Arc[arrow,arrowpos=0.5,arrowlength=5,arrowwidth=5,arrowinset=1,clock](342.667,-8.667)(62.681,150.709,88.781)
    \Arc[arrow,arrowpos=0.5,arrowlength=5,arrowwidth=5,arrowinset=1,clock](377.333,52.667)(62.681,-29.291,-91.219)
    \Vertex(432,22){8}
    \GOval(496,22)(48,16)(0){0.882}
    \Line[arrow,arrowpos=0.5,arrowlength=5,arrowwidth=5,arrowinset=1](512,54)(576,54)
    \Line[arrow,arrowpos=0.5,arrowlength=5,arrowwidth=5,arrowinset=1](576,-10)(512,-10)
    \Text(30,62)[lb]{\Large{\Black{$p+q$}}}
    \Text(125,62)[lb]{\Large{\Black{$k+q$}}}
    \Text(46,-29)[lb]{\Large{\Black{$p$}}}
    \Text(141,-29)[lb]{\Large{\Black{$k$}}}
    \Text(-1,50)[lb]{\Large{\Black{$\alpha$}}}
    \Text(210,50)[lb]{\Large{\Black{$\alpha$}}}
    \Text(365,50)[lb]{\Large{\Black{$\alpha$}}}
    \Text(-1,-15)[lb]{\Large{\Black{$\delta$}}}
    \Text(210,-15)[lb]{\Large{\Black{$\delta$}}}
     \Text(365,-15)[lb]{\Large{\Black{$\delta$}}}
    \Text(184,50)[lb]{\Large{\Black{$\beta$}}}
     \Text(348,50)[lb]{\Large{\Black{$\beta$}}}
      \Text(585,50)[lb]{\Large{\Black{$\beta$}}}
    \Text(184,-15)[lb]{\Large{\Black{$\gamma$}}}
    \Text(348,-15)[lb]{\Large{\Black{$\gamma$}}}
     \Text(585,-15)[lb]{\Large{\Black{$\gamma$}}}
     
     \Text(450,-22)[lb]{\Large{\Black{$r$}}}
     \Text(440,55)[lb]{\Large{\Black{$r+q$}}}
   
    \Arc[arrow,arrowpos=0.5,arrowlength=5,arrowwidth=5,arrowinset=1,clock](364,90)(101.98,-101.31,-138.18)
    \Arc[arrow,arrowpos=0.5,arrowlength=5,arrowwidth=5,arrowinset=1,clock](486.667,52.667)(62.681,-88.781,-150.709)
  \end{picture}
  }
}
\end{center}
Here we have lightened up the picture, removing the superfluous momentum
labels where their placement is clear. The $r$ and $r+q$ internal fermionic
propagators are the full fermionic propagators.

In the case of general momentum $q$ one would consider
the following ansazt for the spinor structure of the vertex
\begin{equation}
V^{\alpha\;\;\;\;\gamma}_{\;\;\delta\;\;,\;\;\beta}(q)=
(\gamma_\mu)^\alpha_{\;\;\delta}
(\gamma_\nu)^\gamma_{\;\;\beta} A^{\mu\nu}(q)\,,
\end{equation}
and therefore the SD equation takes the form
\begin{align}
\label{SDgeneral}
(\gamma_\mu)^\alpha_{\;\;\delta}
(\gamma_\nu)^\gamma_{\;\;\beta} A^{\mu\nu}(q)=\frac{1}{\sigma^2}(\gamma^\mu)^\alpha_{\;\;\delta}
(\gamma_\mu)^\gamma_{\;\;\beta}-\frac{1}{\sigma^2}
(\gamma^\mu)^\alpha_{\;\;\beta}v_{\nu\mu}(q)A^{\nu}_{\;\;\lambda}(q)
(\gamma^\lambda)^\gamma_{\;\;\beta}
\,,
\end{align}
where we have used (\ref{vmunudefinition}).
For the purpose of calculating the Landau parameters
we are interested in the four-fermionic vertex at zero
momentum, $q=0$. In that case
\begin{equation}
\label{vatzeroexchangemomentum}
v_{\mu\nu}(q=0)=\frac{\hat\mu}{4\pi} (\delta _{\mu,1}\delta_{\nu,1}+\delta _{\mu,2}\delta_{\nu,2})
\end{equation}
The SD equation therefore has the solution
\begin{equation}
A_{\mu\nu}(q=0)={\rm diag}\,\left(\frac{1}{\sigma^2+\frac{\hat\mu}{4\pi}},\,\frac{1}{\sigma^2+\frac{\hat\mu}{4\pi}},\,\frac{1}{\sigma^2}\right)\,.
\end{equation}

\subsection{Microscopic derivation of the Landau parameters}

The on-shell particle state $u(p)$ satisfies the Dirac equation
\begin{equation}
\label{DiracEquation}
\hat p_\mu\gamma^\mu u(p)=0\,,
\end{equation}
and therefore it obeys the mass-shell condition
\begin{equation}
(p_3-i\mu-i\Sigma_3)^2+p_s^2=0\,.
\end{equation}
The Euclidean energy is $\varepsilon=\tilde p_3=p_3-i\mu$.
On the Fermi surface the energy is $\varepsilon=\varepsilon_F=-i\mu$, 
and the momentum is $p_F=\hat\mu$.
Dirac equation (\ref{DiracEquation}) on the Fermi surface takes the form
\begin{equation}
\left({-i\atop e^{i\theta_p}}\;{e^{-i\theta_p}\atop i}\right)u(p)=0\,.
\end{equation}
The solution is
\begin{equation}
\label{SolutionU}
u(p)=\left({-ie^{-i\theta_p/2}\atop e^{i\theta_p/2}}\right)\,.
\end{equation}
Now we switch to the Lorentzian signature and expand the fermionic propagator near the Fermi surface
\begin{equation}
G(p)=-\frac{i\hat p_\mu\gamma^\mu}{\hat p^2}\simeq  \frac{1}{\omega -\, (|{\bf p}|-p_F)}\frac{u
\otimes \bar u}{u^\dagger u}\,,
\end{equation}
which has the form of (\ref{LFLpropagator}), with the
wave-function remormalization
\begin{equation}
Z=1\,.
\end{equation}
and the Fermi velocity
\begin{equation}
v_F=1\,.
\end{equation}
Consequently the quasiparticle effective mass is given by
\begin{equation}
\label{meffectiveThirring}
m^\star =\frac{p_F}{v_F}=\hat\mu\,.
\end{equation}

The quasiparticle interaction function $f(\theta)$ can be derived
microscopically from the one-particle-irreducible scattering
amplitude,
(\ref{microscopicf}), (\ref{1PIamplitude}).
Using the solution for $V^{\alpha\;\;\;\;\gamma}_{\;\;\delta\;\;,\;\;\beta}(0)$, and the expression (\ref{SolutionU}) for the on-shell
fermionic state, we obtain
\begin{equation}
f(\theta)=\frac{1}{\sigma^2}-\frac{1}{\sigma^2+\frac{\hat\mu}{4\pi}}\,\cos\theta\,,
\end{equation}
where $\theta=\theta_p-\theta_k$.

The Landau parameters can now be extracted
using the expansion (\ref{fexpansion}).
We expresse the answer in terms of the Thirring
coupling constant $\sigma$ and the Fermi momentum $\hat\mu$,\begin{align}
\label{LandauParameters1}
F_0&=\frac{\hat\mu}{2\pi\sigma^2}\,,\\
\label{LandauParameters2}
F_1&=-\frac{\hat\mu}{4\pi\sigma^2+\hat\mu}\,.
\end{align}

\subsection{Thermodynamic derivation of the Landau parameters}

One can derive the Landau parameters $F_{0}$, $F_1$,
provided the inverse compressibility and the heat capacity are known.
We know that the charge density $n$ is 
given by (\ref{fermiondensity}), which allows us to calculate the
inverse compressibility (\ref{LFLinvsersecompressibility}).
Due to $m^\star=\hat\mu=\mu+\Sigma_3$, we obtain
\begin{align}
\label{F0thermodynamic}
F_0&=N\,\frac{\hat\mu}{2\pi }\frac{\partial \mu}{\partial n}-1=\left(1-\frac{\partial\hat\mu}{\partial\mu}
\right)/\frac{\partial\hat\mu}{\partial\mu}\,.
\end{align}
We notice that this is the same as the expression
(\ref{LandauParameters1}) derived from microscopic considerations,
once we observe that 
\begin{equation}
\left(1-\frac{\partial\hat\mu}{\partial\mu}\right)/\frac{\partial\hat\mu}{\partial\mu}=\frac{\hat\mu}{2\pi\sigma^2}\,,
\end{equation}
as follows from differentiating w.r.t. $\mu$ of the expression,
obtained from equations
(\ref{Sigma3}), (\ref{hatmudefinition})
\begin{equation}
\label{selfenergyrewrite}
\frac{\hat\mu^2}{4\pi}+\sigma^2\hat\mu -\sigma^2\mu=0\,.
\end{equation}

The heat capacity of the Landau Fermi liquid is given by the  expression (\ref{LFLheatcapacity}).
Comparing it with the expression (\ref{heatcapacityThirring}) for the Thirring model
we conclude that $m^\star=\hat\mu$, in agreement
to the value (\ref{meffectiveThirring}) we obtained from
the fermionic propagator. After some transformations using 
the Landau Fermi liquid expression (\ref{LFLeffectivemass}) and the relations (\ref{Sigma3}), (\ref{hatmudefinition})
we conclude that the $F_1$ is given by the expression
(\ref{LandauParameters2}), in agreement with the microscopic derivation of the previous subsection.

\subsection{Solving for the Landau parameters}

\begin{figure}
	\centering
	\begin{minipage}{0.45\textwidth}
	\centering
		\includegraphics[width=1\textwidth]{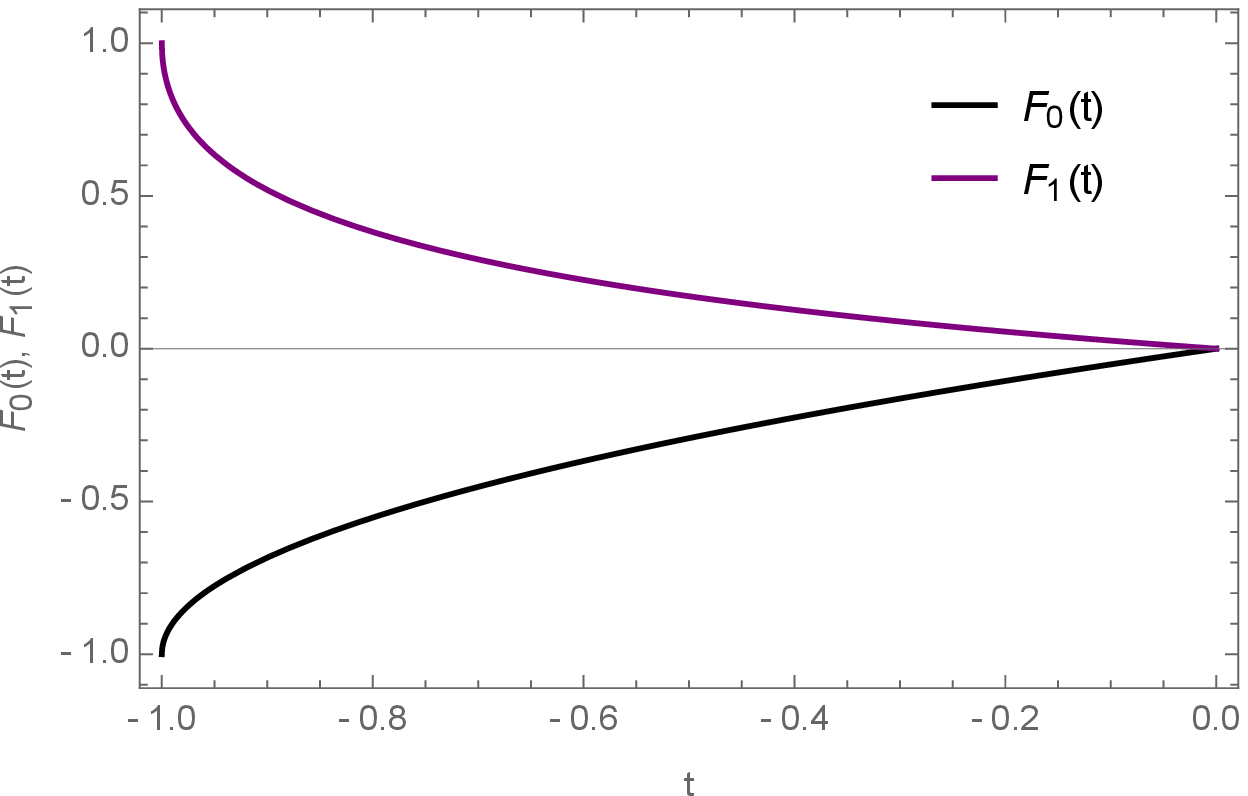}
			\caption{The Landau parameters in the Thirring model with attractive interactions ($\sigma^2<0$) in dependence on the $t=\mu/(\pi\sigma^2)$, taking values between the free point $t=0$ and the threshold point $t=-1$ of the Fermi liquid phase.}
	\label{fig:attractivef}
	\end{minipage}\hfill
	\begin{minipage}{0.45\textwidth}
	\centering
		\includegraphics[width=1\textwidth]{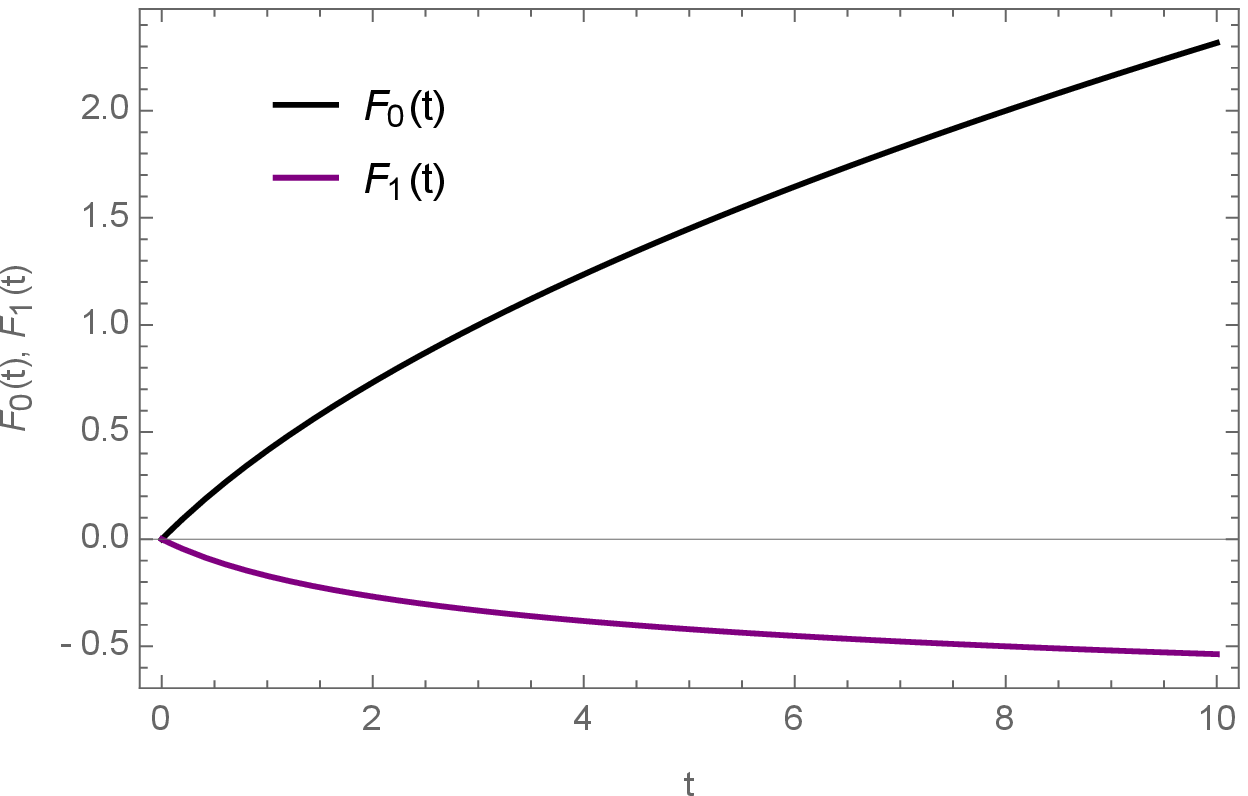}
		\caption{The Landau parameters in the Thirring model with repulsive interactions ($\sigma^2>0$) in dependence on the $t=\mu/(\pi\sigma^2)$, taking values between the free point $t=0$ and the infinitely strongly coupled regime $t\gg 1$.}
		\label{fig:repulsivef}
	\end{minipage}
\end{figure}

The gap equation (\ref{selfenergyrewrite}) is solved by
\begin{equation}
\hat \mu=\mu\,\frac{\sqrt{1+t}-1}{t/2}\,,	
\end{equation}
where we have denoted
\begin{equation}
t=\frac{\mu}{\pi\sigma^2}\,.	
\end{equation}
The fermionic density is then given by
\begin{equation}
	n(t)=N\,\frac{\mu^2}{\pi\, t^2}\,(\sqrt{1+t}-1)^2\,.
\end{equation}
The Landau parameters are
\begin{align}
F_0(t)&=\sqrt{1+t}-1\,,\\
F_1(t)&=-\frac{1}{t}	\,(\sqrt{1+t}-1)^2\,.
\end{align}
We plot these for the attractive and repulsive interactions in
figures \ref{fig:attractivef}, \ref{fig:repulsivef}. 

The free theory limit is $t\rightarrow 0$. In the case of attractive
interaction we have $t<0$, in the case of repulsive interaction it is $t>0$.
The theory always has a real-valued solution for $\hat\mu$
when interaction is repulsive. When interaction is attractive, the solution is
only valid for $t\in [-1,0]$, putting an upper threshold $t=-1$ on the
possible interaction strength at which the Fermi liquid state can exist.
Notice the density $n(t)$ does not vanish at the threshold point $t=-1$, indicating that the system actually goes through a phase transition at this point, to a different finite-density state.
This is not unexpected for a three-dimensional model with local four-fermionic interaction, since it is known that a finite-density
Gross-Neveu model exhibits a deconfinement phase transition at
certain value of the chemical potential \cite{Rosenstein:1988dj,Rosenstein:1990nm}. It would be interesting to derive the free energy for the Thirring model, at finite temperature and chemical potential, and map the corresponding phase struture.

\section{Discussion}\label{sec:disc}

In this paper we considered the large $N$ limit of the three-dimensional Thirring model of massless fermions at finite density. Such a model provides a simple example of an interacting fermionic system, and we have shown that its low-temperature dynamics consistently fits into the framework of the Landau Fermi liquid with a non-trivial zeroth and first Landau parameters. We have argued that the system exhibits a sharp Fermi surface at zero temperature by calculating the current-current correlation function and demonstrating that it has a singular structure at zero frequency and a finite momentum, equal to twice of the Fermi momentum.

An immediate generalization of the model considered
in this paper is achieved by switching on a finite temperature.
The Landau Fermi liquid theory is defined in the low-temperature range, $T/\mu\ll 1$.
Unlike the situation of the Chern-Simons-matter theories, introduction of temperature into
the Thirring model is relatively simple, because in the absence of gauge interaction
one does not have to worry about holonomy of the gauge field along the temporal circle. It is  interesting to calculate the current-current correlation
function at finite value of the temperature, and study its singular structure.

Another straightforward calculation which can be done
is the large-$N$ free energy and the associated phase structure.
Furthermore, it would be interesting to repeat the
analyses of this paper for the three-dimensional massive fermionic model with the Gross-Neveu interaction, which exhibits a second order phase transition \cite{Rosenstein:1988dj,Rosenstein:1990nm}.
It would be interesting to observe the associated behavior of the Landau parameters and to follow
the fate of the Friedel oscillations across the point of the superconducting phase transition.

\section*{Acknowledgements}
This work was supported by the Oehme Fellowship.
I would like to thank B.~Galilo, M.~Geracie and M.~Roberts for useful discussions.
I would like to thank Technion-Israel Institute of Technology, where part of this
work was completed, for hospitality.

\appendix

\section{Derivation of the $v$ matrix}\label{App:A}

In this appendix we derive the $v_{\mu\nu}$ matrix (\ref{vmunudefinition})
\begin{equation}
v_{\mu\nu}={\rm Tr}(\gamma_\mu\gamma_\alpha\gamma_\nu
\gamma_\beta)\,\int\frac{d^3r}{(2\pi)^3}\frac{1}{\hat r^2(\hat r+q)^2}\hat r^\alpha (
\hat r+q)^\beta\,.
\end{equation}
Let us introduce the $u_{\alpha\beta}$ matrix
\begin{equation}
u^{\alpha\beta}=\int\frac{d^3r}{(2\pi)^3}\frac{1}{\hat r^2(\hat r+q)^2}\hat r^\alpha (
\hat r+q)^\beta\,,
\end{equation}
which then allows us to express
\begin{equation}
v_{\mu\nu}=2(u_{\mu\nu}+u_{\nu\mu}-{\rm Tr}\, u\, g_{\mu\nu})\,.
\end{equation}
We are interested in the calculation at $q_3=0$.
The calculation therefore reduces to deriving the following integrals:
\begin{align}
S_\mu(q_3=0)&=\int\frac{d^3r}{(2\pi)^3}\frac{\hat r_\mu}{\hat r^2 (\hat r+q)^2}\,,\\
S_{\mu\nu}(q_3=0)&=\int\frac{d^3r}{(2\pi)^3}\frac{\hat r_\mu
\hat r_\nu}{\hat r^2 (\hat r+q)^2}\,,
\end{align}
in terms of which
\begin{equation}
u^{\alpha\beta}(q_3=0)=S^{\alpha\beta}(q_3=0)+S^\alpha (q_3=0)q^\beta\,.
\end{equation}
For simplicity of notation in this appendix we omit hats
on top of $\hat\mu$.

Let us derive the $S_\mu$ first. Introducing the Feynman parameter
we obtain
\begin{align}
S_\mu(q_3=0)=\int _0^1 dx\, \int \frac{d^3 r}{(2\pi)^3}\,\frac{\tilde r_\mu}{[\tilde r_3^2+a^2]^2}\,,
\end{align}
where $b^2=(r+xq)_s^2+x(1-x)q_s^2$. First integrating
over $r_3$ we notice that
\begin{equation}
S_3(q_3=0)=0\,.	
\end{equation}
Consider $\mu=i$
to be a spatial polarization, and make the change 
$r_i\rightarrow r_i -x\,q_i$ of the integrated momentum,
\begin{align}
S_i(q_3=0)&=-q_i\,\int _0^1 dx\, x\, \int \frac{d^3 r}{(2\pi)^3}\,\frac{ 1}{[\tilde r_3^2+a^2]^2}\\
&=-\frac{q_i}{8\pi}\,\int _0^1 dx\, x\, \int \frac{d^2 r}{(2\pi)^2}\,
\int _{\sqrt{x(1-x)}q_s}^{\Lambda+x q_s \cos\theta}\frac{da}{a^2}\,\theta (a-\mu)\,,
\end{align}
where $a^2=r_s^2+x(1-x)q_s^2$.

Here we have also noticed that on should generally take into account that if the $r_s$
integral is divergent then an extra subtlety appears in
regularization of the divergence. Suppose we choose a cutoff
scale, $r_s<\Lambda$. Then after the change of the variables
the cutoff scale is $a<\Lambda+xq_s\cos\theta+{\cal O}(1/\Lambda)$,
where $\theta$ is an angle between ${\bf r}$ and ${\bf q}$.
Due to the presence of $\cos\theta$, the shift $xq_s\cos\theta$ of
the cutoff scale usually vanishes after integration over $\theta$,
and in all the integrals below it actually ends up having no contribution.

Denote $x_{1,2}$ to be solutions of equation
$x(1-x)q_s^2=\mu^2$. Then
\begin{align}
S_i(q_3=0)&=-\frac{q_i}{8\pi}\,\int _0^1 dx\, x\,\int _\mu\frac{da}{a^2}\,\theta(2\mu-q_s)\\
&-\frac{q_i}{8\pi}\left(\left(\int _0^{x_1}dx\,x+\int _{x_2}^1dx\,x\right)\frac{1}{\mu}+\int _{x_1}^{x_2}dx\,x\frac{1}{q_s\sqrt{x(1-x)}}\right)\,\theta(q_s-2\mu)\notag\\
&=-\frac{q_i}{16\pi\mu}+\frac{1}{8\pi}	\left(
\frac{\sqrt{q_s^2-4\mu^2}}{2\mu}-\cos^{-1}\left(\frac{2\mu}{q_s}\right)
\right)\,\theta(q_s-2\mu)\,.
\end{align}
Therefore for the polarizations longitudinal and transverse w.r.t. ${\bf q}$ we obtain
\begin{align}
	S_{\parallel}(q_3=0)&=-\frac{q_s}{16\pi\mu}+\frac{1}{16\pi\mu}	\left(
\sqrt{q_s^2-4\mu^2}-2\mu\,\cos^{-1}\left(\frac{2\mu}{q_s}\right)
\right)\,\theta(q_s-2\mu)\,,\\
S_{\perp}(q_3=0)&=0\,.
\end{align}

Calculation of the tensor $S_{\mu\nu}$ is performed analogously. 
First of all we notice that
\begin{align}
S_{3i}(q_3=0)&=0\\
S_{33}(q_3=0)&=-\frac{\mu}{8\pi}+\left(\frac{\mu}{16\pi q_s}\sqrt{q_s^2-4\mu^2}
-\frac{q_s}{32\pi}\cos^{-1}\frac{2\mu}{q_s}\right)\theta (q_s-2\mu)\,.
\end{align}
We subseqeuntly find
\begin{align}
S_{ij}(q_3=0)&=\frac{1}{8\pi}\,\int _0^1 dx\int\frac{d\theta}{2\pi}\int_{\sqrt{x(1-x)}q_s}^{\Lambda+ x q_s\cos\theta} \frac{da}{a^2}\,\theta(a-\mu)\,	(r_ir_j+x^2q_iq_j)\,.
\end{align}
which in components is given by
\begin{align}
S_{\parallel\parallel}(q_3=0)&=\frac{q_s^2-2 \mu ^2}{32 \pi  \mu }+\frac{2 \mu  q_s \cos ^{-1}\left(\frac{2 \mu }{q_s}\right)-q_s \sqrt{q_s^2-4 \mu ^2}}{32 \pi  \mu }\,\theta (q_s-2\mu)\,,\\
S_{\perp\perp}(q_3=0)&=-\frac{q_s^2+6 \mu ^2}{96 \pi  \mu }+
\frac{\sqrt{q_s^2-4 \mu ^2 } \left(8 \mu ^2+q_s^2-6 \mu ^2 \cos ^{-1}\left(\frac{2 \mu }{q_s}\right)\right)}{96 \pi  \mu  q_s}\,
\theta(q_s-2\mu)\\
S_{\parallel\perp}(q_3=0)&=0\,.
\end{align}

Notice that all the integrals are real-valued at $q_3=0$.
This can alternatively be seen by working in the Lorentzian signature, where the fermionic propagator at finite chemical potential attains the form
\begin{equation}
\frac{1}{-p_3^2+p_s^2}\rightarrow \frac{1}{(p_3+p_s-i\epsilon)
(p_3-p_s+i\epsilon {\rm sgn}(p_s-\hat\mu))}\,.
\end{equation}
Using this propagator in the loop integral at zero frequency one notices that the resulting integrals over spatial components of the total momentum are to be understood in the principal value sense, and can be seen to be real-valued.

For the density-density correlation function, restoring $\mu\rightarrow\hat\mu$, we obtain
\begin{align}
&\frac{1}{N}\,\left\langle j_3(-q)j_3(q)\right\rangle=\frac{q_s^2 \sigma ^2\,\theta (2\hat\mu -q_s)}{12 \pi  \hat\mu  \sigma ^2+q_s^2}	\\
&{+}\frac{\sigma ^2 \left(4 q_s^3{-}4 q_s^2 \sqrt{q_s^2{-}4 \hat\mu ^2}{-}2 \hat\mu ^2 \sqrt{q_s^2{-}4 \hat\mu ^2}+3 \hat\mu  \left(2 \hat\mu  \sqrt{q_s^2{-}4 \hat\mu ^2}{+}q_s^2\right) \cos ^{-1}\left(\frac{2 \hat\mu }{q_s}\right)\right)\,\theta (q_s{-}2\hat\mu)}{4 q_s^3{-}4 q_s^2 \sqrt{q_s^2{-}4 \hat\mu ^2}{-}2\hat\mu ^2 \sqrt{q_s^2{-}4 \hat\mu ^2}{+}3 \hat\mu  \left(2 \hat\mu  \sqrt{q_s^2{-}4 \hat\mu ^2}{+}q_s^2\right) \cos ^{-1}\left(\frac{2 \hat\mu }{q_s}\right){+}48 \pi  \hat\mu  q_s \sigma ^2}\,.\notag
\end{align}

\bibliographystyle{JHEP}
\bibliography{MGbib}

\end{document}